\documentclass{article}
\usepackage{spconf,amsmath,graphicx}
 \usepackage{graphicx}
 \if CLASSOPTIONcompsoc
 \usepackage[caption=false, font=normalsize, labelfont=sf, textfont=sf] {subfig}
 \else
 \usepackage[caption=false, font=footnotesize]{subfig}
\fi
\usepackage{url}
\usepackage{multirow}
\usepackage{booktabs}
\usepackage{balance}
\usepackage{amssymb}

\usepackage{glossaries}
\newacronym{cdf}{CDF}{cumulative distribution function}
\newacronym{pdf}{PDF}{probability density function}
\newacronym{mse}{MSE}{mean-squared error}
\newacronym{bs}{BS}{base station}
\newacronym{rssi}{RSSI}{received signal strength indicator}
\newacronym{nlos}{NLOS}{non-line-of-sight}
\newacronym{los}{LOS}{line-of-sight}
\newacronym{sfo}{SFO}{sampling frequency offset}
\newacronym{cfo}{CFO}{carrier frequency offset}
\newacronym{pbd}{PBD}{packet boundary detection}
\newacronym{csi}{CSI}{channel state information}
\newacronym{cir}{CIR}{channel impulse response}
\newacronym{ofdm}{OFDM}{orthogonal frequency-division multiplexing}
\newacronym{awgn}{AWGN}{additive white Gaussian noise}
\newacronym{agc}{AGC}{automatic gain control}
\newacronym{boi}{BoI}{band-of-interest}
\newacronym{mad}{MAD}{mean absolute deviation}
\newacronym{psd}{PSD}{power spectral density}
\newacronym{dw}{DW}{discrete wavelet}
\newacronym{dft}{DFT}{discrete Fourier transform}
\newacronym{idft}{IDFT}{inverse discrete Fourier transform}

\newacronym{pll}{PLL}{phase-locked loop}

\newacronym{rfid}{RFID}{radio-frequency identification}
\newacronym{ble}{BLE}{Bluetooth low energy}
\newacronym{rss}{RSS}{received signal strength}
\newacronym{cfr}{CFR}{channel frequency response}
\newacronym{toa}{ToA}{time-of-arrival}
\newacronym{aoa}{AoA}{angle-of-arrival}
\newacronym{uwb}{UWB}{ultra-wideband}

\newacronym{kld}{KLD}{Kullback-Leibler divergence}
\newacronym{gmm}{GMM}{Gaussian mixture model}
\newacronym{mdgmm}{MD-GMM}{multi-dimensional Gaussian mixture model}
\newacronym{twr}{TWR}{two-way ranging}
\newacronym{tdoa}{TDoA}{time-difference-of-arrival}
\newacronym{cnc}{CNC}{computer numerical control}

\newacronym{svc}{SVC}{support vector classification}

\title{Single-anchor UWB Localization using \\ Channel Impulse Response Distributions
 }
\name{Sitian Li$^{\star}$ \qquad Alexios Balatsoukas-Stimming$^{\dagger}$ \qquad Andreas Burg$^{\star}$\thanks{This research has been kindly supported by the Swiss National Science Foundation under Grant-ID 182621.}}
\address{$^{\star}$ Telecommunication Circuits Laboratory, \'{E}cole polytechnique f\'{e}d\'{e}rale de Lausanne, Switzerland \\
$^{\dagger}$ Eindhoven University of Technology, The Netherlands}

\begin{document}
\ninept

\maketitle

\begin{abstract}
Ultra-wideband (UWB) devices are widely used in indoor localization scenarios.
Single-anchor UWB localization shows advantages because of its simple system setup compared to conventional two-way ranging (TWR) and trilateration localization methods.
In this work, we focus on single-anchor UWB localization methods that learn statistical features of the channel impulse response (CIR) in different location areas using a Gaussian mixture model (GMM). 
We show that by learning the joint distributions of the amplitudes of different delay components, we achieve a more accurate location estimate compared to considering each delay bin independently.
Moreover, we develop a similarity metric between sets of CIRs.
With this set-based similarity metric, we can further improve the estimation performance, compared to treating each snapshot separately. 
We showcase the advantages of the proposed methods in multiple application scenarios. 
\end{abstract}
\begin{keywords}
Single-anchor localization, channel impulse response, ultra-wideband, Gaussian mixture model
\end{keywords}
\section{Introduction}
Indoor localization using wireless communications signals has a wide range of applications, such as asset tracking, occupancy analytics, and navigation.
Due to its low power consumption, short pulse duration, and robustness against multipath fading, \gls{uwb} has become a popular indoor localization method~\cite{xiao_survey_2010}.
The IEEE 802.15.4a-2007 standard~\cite{noauthor_ieee_nodate} first proposed \gls{twr}, which enables measuring the time of the signal transmission between two devices, which are usually referred to as the tag and the anchor, in the absence of clock synchronization, in order to calculate the distance between the anchor and the tag.
To estimate the location, a \gls{twr} system consisting of at least three anchors at known positions is required.
The location of the tag is estimated by the distance between the tag and the anchors using trilateration.
However, \gls{twr} still relies on devices with extremely stable clocks~\cite{mikhaylov_selection_2016, sidorenko_decawave_2019} and requires \gls{los} transmission between the tag and anchors to operate well~\cite{heydariaan_toward_2018}.

Single-anchor \gls{uwb} localization methods have attracted significant attention recently. 
Instead of regarding the multipath channel as a problem for positioning as in \gls{twr}, single-anchor methods regard reflected transmission paths as virtual anchors~\cite{groswindhager_uwb-based_2017, groswindhager_salma_2018}.
A very important channel characteristic, namely the \gls{cir}, is a vector containing multiple delay bins calculated by the receiver.
Each delay bin is a superposition of multiple transmission paths with different delays in the physical environment.
For each possible position of the anchor, the expected \gls{cir} from the anchor and virtual anchors are calculated and compared with the measured \gls{cir}.
However, in order to know the accurate position of the anchors to localize the tag, the structure of the room (floor plan) needs to be obtained.
Furthermore, only a few reflection paths on the walls have been considered in the literature, and high-order reflections, diffraction or diffuse scattering are not considered, which limits the positioning accuracy.

Since it is almost impossible to have a closed-form expression of the \gls{cir} in a room with a complex structure, learning-based approaches have been proposed to identify reasonably accurate mappings between the \gls{cir} and the position of the tag.
Several works train a localization model to take a \gls{cir} snapshot, the linear projection of a \gls{cir} onto a low dimensional space, and other features such as the \gls{rssi} as input to the model~\cite{rana_uwb_2017,krishnan_improving_2018, lu_deep_2021}.
However, there are still drawbacks to training from individual \gls{cir} snapshots. 
First, due to the lack of synchronization between the tag and the anchor~\cite{hol_ultra-wideband_2010, zhuo_identifying_2016, xie_precise_2019}, the measured \gls{cir} is shifted randomly and is multiplied with an additional random phase term. 
Even if the tag is stable, the received \gls{cir}s can be different since there is always a residual  offset after the packet detection.
Second, since the resolution of \gls{uwb} is not sufficient to retrieve and separate all reflected paths, a \gls{cir} snapshot is a low-dimensional representation of a high-dimensional physical environment.
Hence, we potentially need more snapshots as additional information for a good location estimate.

Therefore, the work of~\cite{mohammadmoradi_uwb-based_2019} proposed to consider \gls{cir} statistics instead of individual realizations.
One intuitive method to learn a statistic is to assume a \gls{gmm}, which has been applied successfully to other wireless communication positioning methods such as finger-printing in WiFi localization~\cite{shi_probabilistic_2016}.
In those methods, the room is first divided into several pre-defined areas (around $0.01$\,m$^2$), a large number of \gls{cir}s are sampled in each area, and the distribution for each delay bin in the \gls{cir} in each area is recorded and modeled by a \gls{gmm}.
When a new \gls{cir} observation becomes available, a log-probability-based score is calculated for each reference area and the area with the highest score is the location estimate.
The score is calculated by treating the delay bins in one \gls{cir} as independent variables.
However, better performance can be expected when considering one \gls{cir} with multiple delay bins as a sample from a multi-variate distribution of the pre-defined areas.
Moreover, when the pre-defined areas become larger, the true physical distribution becomes challenging to model with a limited number of Gaussian distributions, and it becomes difficult to distinguish between the pre-defined areas. 
In~\cite{mohammadmoradi_uwb-based_2019}, majority voting is proposed, which considers the hard decisions from multiple \gls{cir} snapshots during movement in the same area and improves the estimation accuracy. 
Since multiple samples are needed for the majority voting, a joint decision on the soft scores is expected to perform better than majority voting among individual hard decisions.
\vspace{-0.3cm}
\subsection*{\small{Contributions}}
In this work, instead of assuming that \gls{cir} bins are independent, we use a multi-bin \gls{gmm}-based joint distribution.
Moreover, we consider multiple \gls{cir} snapshots as a new sample set and propose to define a joint decision-based similarity metric between the new sample set and the sample sets in reference areas.
To localize a set of \gls{cir}s with the similarity metric, we introduce two estimation methods:
One is to simply choose the candidate with the maximum similarity, and the other is to apply machine learning to obtain the mapping between vectors of similarity metrics and the location area of the samples.
Single-anchor localization, on the one hand, relies on a rich set of multipaths to estimate the position, but on the other hand, can be sensitive to multipath component changes, which decreases the localization accuracy. 
We thus test the performance of several methods in different environments, whose layout complexity varies.
Our methods improve the estimation accuracy by up to $40\%$ compared to the methods proposed until now in the literature.
We also explore the robustness of single-anchor localization against layout changes in the same room.

\vspace{-0.2cm}
\section{Background}
In this section, we first briefly introduce the \gls{cir} and how it is estimated in the \gls{uwb} receiver. 
Then, the localization method from~\cite{mohammadmoradi_uwb-based_2019} is introduced as a baseline.
\vspace{-0.3cm}
\subsection{Channel Impulse Response (CIR)}
The IEEE 802.15.4 UWB preamble sequences~\cite{noauthor_ieee_nodate} have a perfect periodic auto-correlation, which in essence allows a coherent receiver to estimate the \gls{cir} between the transmitter and the receiver~\cite{decawave_dw1000_nodate_um}.
There is usually a distortion residual in the estimated \gls{cir} because of the non-synchronized transmitter and receiver.
In the following, the estimated \gls{cir} is denoted by a complex-valued vector $\mathbf{h}$.
\vspace{-0.3cm}
\subsection{One-dimensional Gaussian Mixture Model (1D-GMM)}
\label{sec:baseline}
The localization method described in~\cite{mohammadmoradi_uwb-based_2019} is divided into two phases.
In the offline learning phase, a large number of \gls{cir}s $\mathbf{h}_c^t$ are measured from each pre-defined area $c \in \mathcal{C}$, where $t$ is the snapshot index and $\mathcal{C}$ denotes the set of reference area indices.
Each $\mathbf{h}_c^t$ is a vector whose entries represent delay bins of the \gls{cir}.
For each delay bin~$m$ and area $c$, the amplitude of $\mathbf{h}_c^t[m]$ for all snapshot indices $t$ is fitted using a \gls{gmm} whose \gls{pdf} is denoted by $g_c^m$. 
In the online phase, the estimated area $\hat{l}^t$ for a new \gls{cir} sample $\mathbf{h}^t$ is given by
\vspace{-0.1cm}
\begin{align}
\hat{l}^t = \arg \max_{c\in \mathcal{C}} \sum_{m=0}^{M-1} \log g_c^m(|\mathbf{h}^t[m]|),
\label{equ:gmm_baseline}
\end{align}
where $M$ is the number of considered delay bins.
\section{Proposed Methods}
We first propose our multi-dimensional \gls{gmm} method.
Then, we describe the position estimation method using the similarity metric.
\label{sec:methods}
\vspace{-0.2cm}
\subsection{Multi-dimensional Gaussian Mixture Model (MD-GMM)}
\begin{figure}[t!]
	\centering
	\includegraphics[scale=0.5]{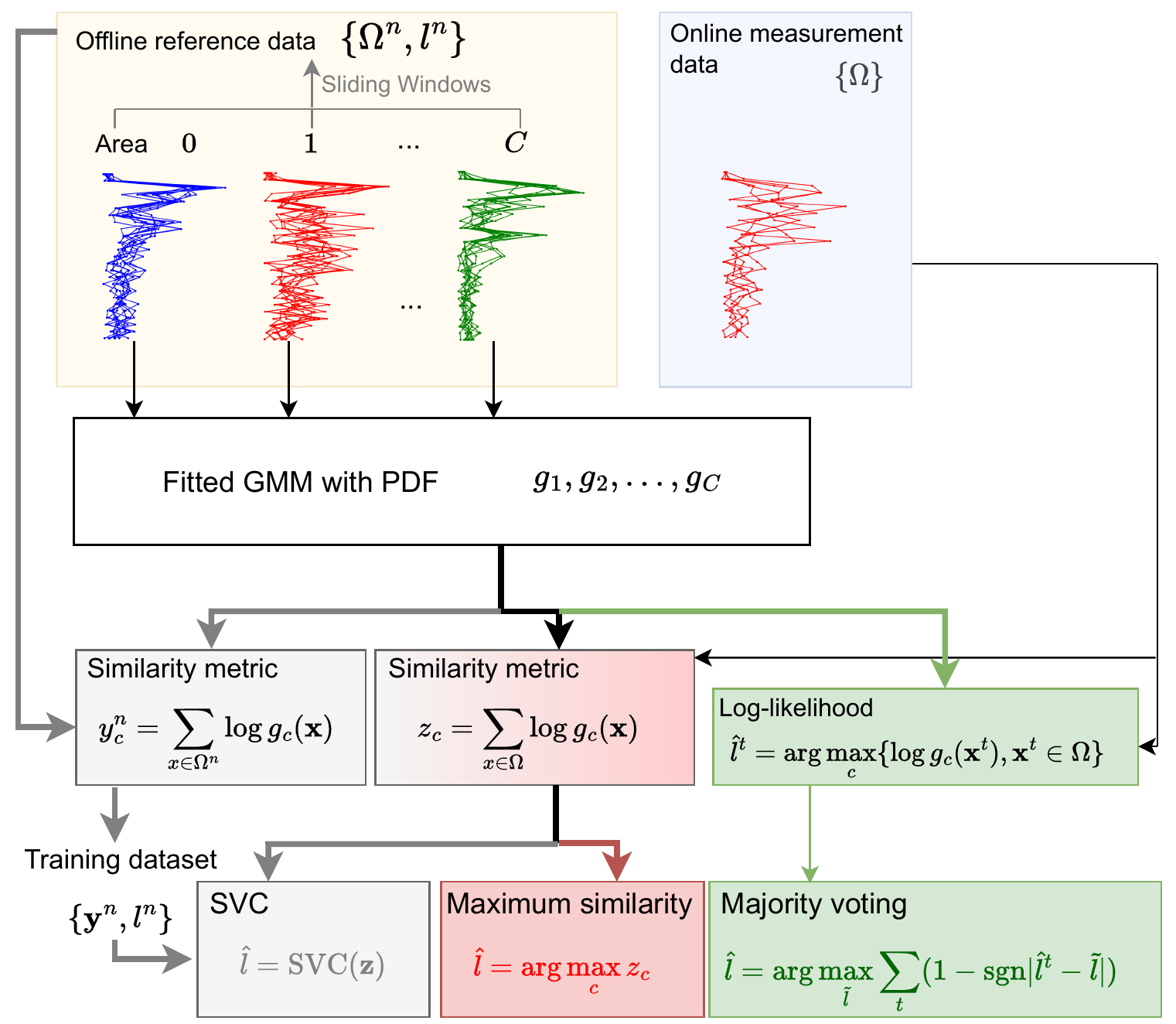}
	\caption{Block Diagram of MD-GMM (green), MD-GMM-MaxSim (red) and MD-GMM-SVC (gray). The yellow box denotes reference CIR set, the blue box denotes one sample set during online measurement phase. The white box is the GMM fitting block that is shared by all methods.}\label{fig:bd}
	\vspace{-0.1cm}
\end{figure}
Note that (\ref{equ:gmm_baseline}) treats the magnitudes $|\mathbf{h}^t[m]|$ at multiple delay bins $m$ as independent variables. 
However, because of the potentially complex reflection environment, the adjacent delay bins can be impacted by the same delay path and are thus generally correlated.
For this reason, we propose to consider each $|\mathbf{h}^t|$ as a multi-variate random variable and to fit a $M$-dimensional \gls{gmm} whose \gls{pdf} is $g_c$ for each area $c$ in the offline training phase, as shown by the white box in Fig.~\ref{fig:bd}. 
In the online test phase, we measure the log-probability of $\mathbf{h}^t$ for each $t$ being sampled from all the individual reference \gls{gmm}s.
The estimated area $\hat{l}^t$ for $\mathbf{h}^t$ is then defined as
\begin{align}
\hat{l}^t = \arg \max_{c \in \mathcal{C}} \log g_c(|\mathbf{h}^t|).
\end{align}
Intuitively, by inspecting all delay bins of the \gls{cir} jointly, the performance should benefit from considering the multi-variate \gls{cir} distribution.
To further improve the accuracy, we then apply a majority vote as in~\cite{mohammadmoradi_uwb-based_2019} according to
\begin{align}
\hat{l} = \arg \max_{\tilde{l} \in \mathcal{C}} \sum_{t=1}^{T} (1-\text{sgn}(|\hat{l}^t - \tilde{l}|)),
\end{align}
for a set of consecutive samples $|\mathbf{h}^t|$, denoted by $\Omega$, where $\text{sgn}(\cdot)$ is the sign function and $T$ is the number of snapshots.
\begin{figure}[t!] 
\vspace{-0.6cm}
    \centering
  \subfloat[Small-scale bin 7\label{fig:s7}]{%
        \includegraphics[width=0.48\linewidth]{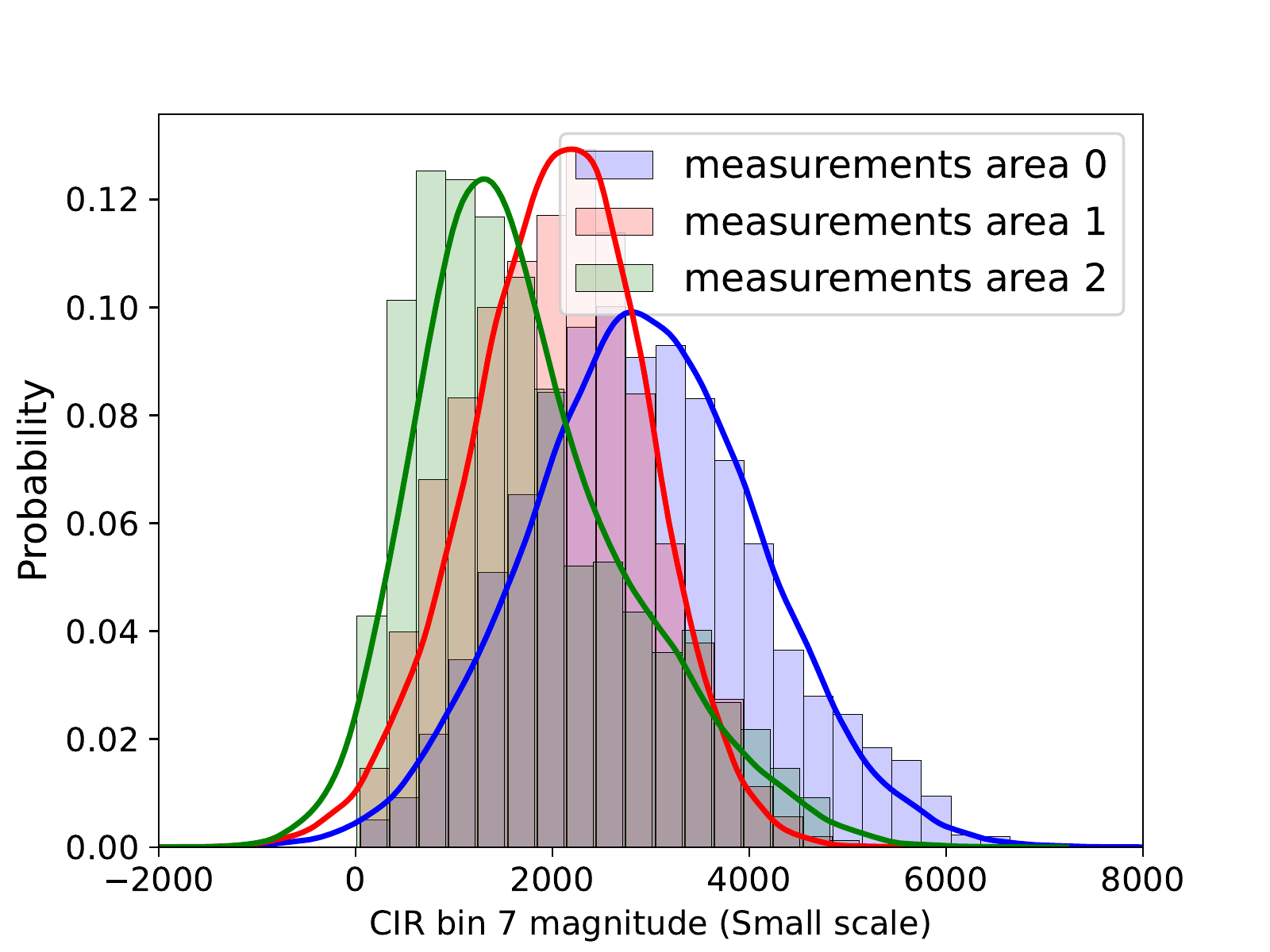}}
    \hfill
  \subfloat[Large-scale (Computer room NLOS) bin 7\label{fig:l7}]{%
        \includegraphics[width=0.48\linewidth]{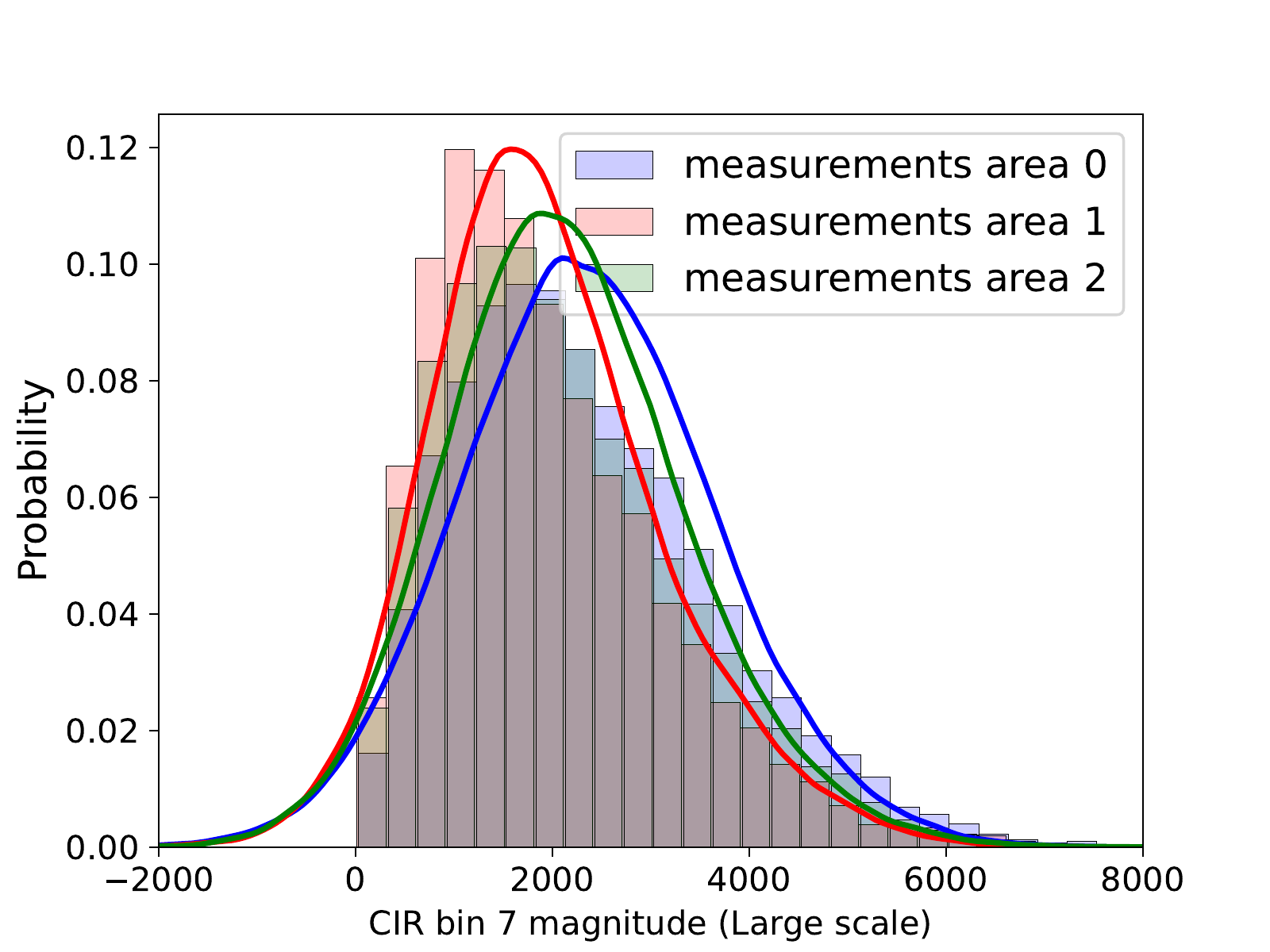}}
  \caption{Marginal CIR magnitude histograms (rectangles) and fitted MD-GMM marginal distributions (solid lines) in different areas}
  \label{fig:l} 
  \vspace{-0.5cm}
\end{figure}
\vspace{-0.3cm}
\subsection{Similarity Metric based on MD-GMM}
Even though the performance can be improved by majority voting from multiple snapshots, there are two drawbacks to the previously described log-probability-based methods: 
First, we observed that when the distribution of the physical channel is complicated, the \gls{pdf}s become difficult to distinguish between areas, especially when the area in each location is large (e.g., $1$\,m$^2$) and the environment has many obstacles.
As shown in Fig.~\ref{fig:l}, we plot the marginal \gls{cir} distributions on delay bin $7$. 
For the \gls{nlos} case, the distributions and the fitted \gls{gmm}s of different areas can be more similar compared to the small-scale case.
Second, the majority voting based on hard decisions can be improved by a joint soft decision among multiple snapshots.
Thus we propose to regard multiple snapshots as samples from a new distribution $\Omega$, and estimate the location by a similarity metric between the distribution of a new sample set and the distributions of the reference sample sets in different areas.

The estimation based on the log-probability with the \gls{gmm} is in fact closely related to searching for the reference area $\hat{l}$ whose \gls{gmm} with the \gls{pdf} $g_{\hat{l}}$ has the smallest \gls{kld} with the sampled set $\Omega$.
More specifically, suppose that the \gls{pdf} of $\Omega$ is $f_{\Omega}(\mathbf{x})$.
Then we have
\begin{align}
\hat{l} &= \arg\min_{c\in\mathcal{C}} \text{KLD}(f_{\Omega}\|g_c) = \arg\min_{c\in\mathcal{C}} \int f_{\Omega}(\mathbf{x}) \log \frac{f_{\Omega}(\mathbf{x})}{g_c(\mathbf{x})} d \mathbf{x} \\
&= \arg\max_{c\in\mathcal{C}} \int f_{\Omega}(\mathbf{x}) \log g_c(\mathbf{x}) d \mathbf{x}  \\
&\approx \arg\max_{c\in\mathcal{C}} \sum_{t=1}^{T} \log g_c(\mathbf{x}^t), \label{equ:approxi}
\end{align}
where $\mathbf{x}^t$, $t = 1\hdots T$ are i.i.d. sampled from the distribution $f(\mathbf{x})$. 
Eq.~(\ref{equ:approxi}) becomes an equality when $T \rightarrow \infty$.
Inspired by this observation, we define
\vspace{-0.25cm}
\begin{align}
z_c = \sum_{t=1}^{T} \log g_c(|\mathbf{h}^t|)
\label{equ:similarity_metric}
\end{align}
as the similarity metric between $\Omega$ and $g_{c}$ and we pick the class $c$ with the largest similarity measure as the location estimate.
We refer to this method illustrated by red boxes in Fig.~\ref{fig:bd} as MD-GMM-MaxSim.

The similarity metric is defined based on the assumption that all samples from $\Omega$ are i.i.d. distributed, which is not true since the samples are typically recorded over a short time window.
As an alternative method, we propose to learn the mapping between the location area and the similarity metric vector $\mathbf{z}$. 
We call this method MD-GMM-SVC and we use reference datasets for training as well as a \gls{svc} as shown with gray boxes in Fig.~\ref{fig:bd}.
Compared to other methods, MD-GMM-SVC has an additional training step after the joint \gls{gmm}s are obtained.
The reference dataset is sliced into sliding windows indexed by $n$ of $T$ snapshots each, which creates sample sets of $|\mathbf{h}|$, denoted by $\Omega^n$, with corresponding labels $l^n$ indicating the area in which the sample sets are collected.
We compute the similarity metric as in (\ref{equ:similarity_metric}).
The set of similarity metrics $\mathbf{y}^n$ together with the labels $l^n$ are used to learn the mapping between the similarity metric and the location area using \gls{svc}.
In the online test phase, for a set of new observations $\Omega$, we first measure the similarity metric $\mathbf{z}$, and then use the trained \gls{svc} model to decide the location area $\hat{l}$.

\vspace{-0.1cm}
\section{Experimental Study}
In this section, we evaluate the performance of the three methods described in Section~\ref{sec:methods} in different scenarios with the method in Section~\ref{sec:baseline} as the baseline.
For each scenario, we divide the considered area into $12$ sub-areas, and each has an area label.
We localize the transmitter by deciding in which area the transmitter is positioned.
\begin{figure}
    \centering
  \subfloat[Classroom layout\label{fig:classroom}]{%
       \includegraphics[width=0.43\linewidth]{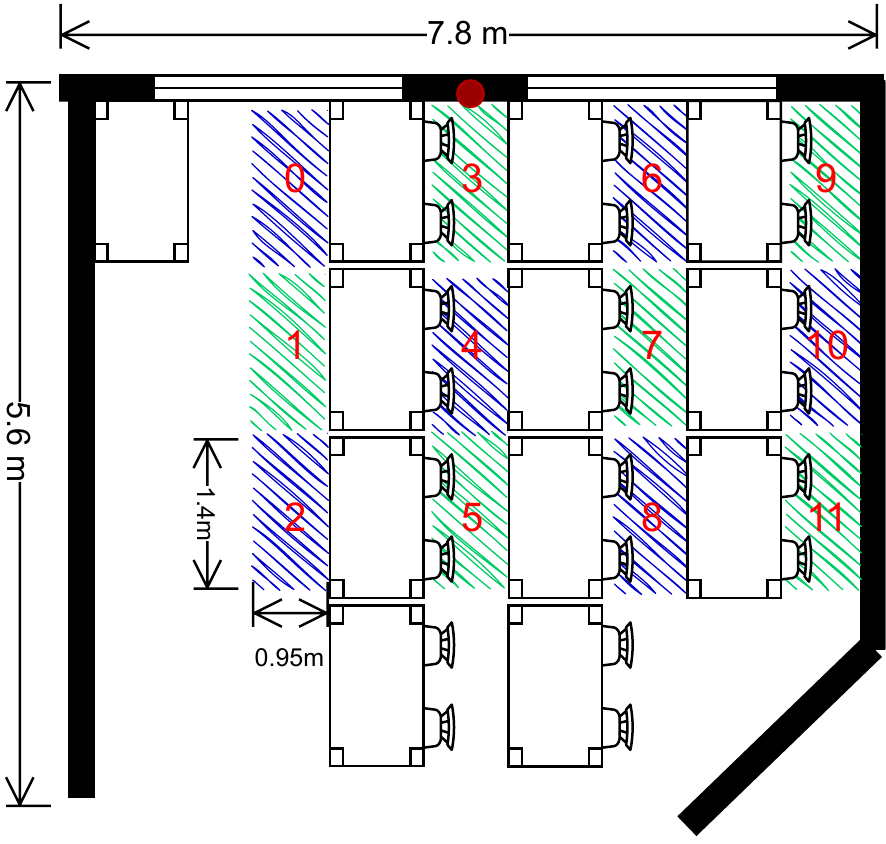}}
    \hfill
  \subfloat[Computer room layout \label{fig:labroom}]{%
        \includegraphics[width=0.5\linewidth]{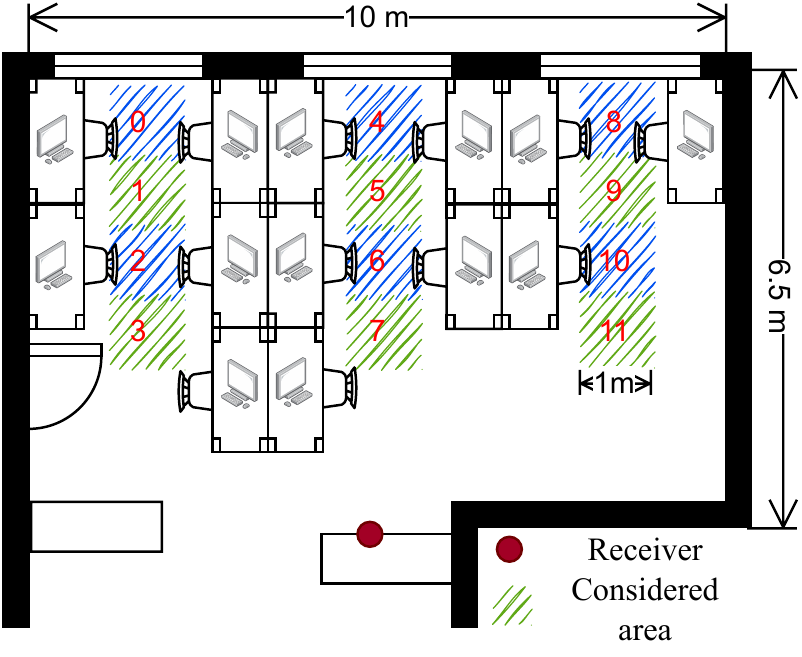}}
  \caption{Layout of the large-scale setups}
  \label{fig:rooms} 
  \vspace{-0.4cm}
\end{figure}
\begin{figure} [t!]
    \centering
  \subfloat[Original classroom layout (top) and layout with moved chairs (bottom).\label{fig:classroom_photo}]{%
       \includegraphics[width=0.47\linewidth]{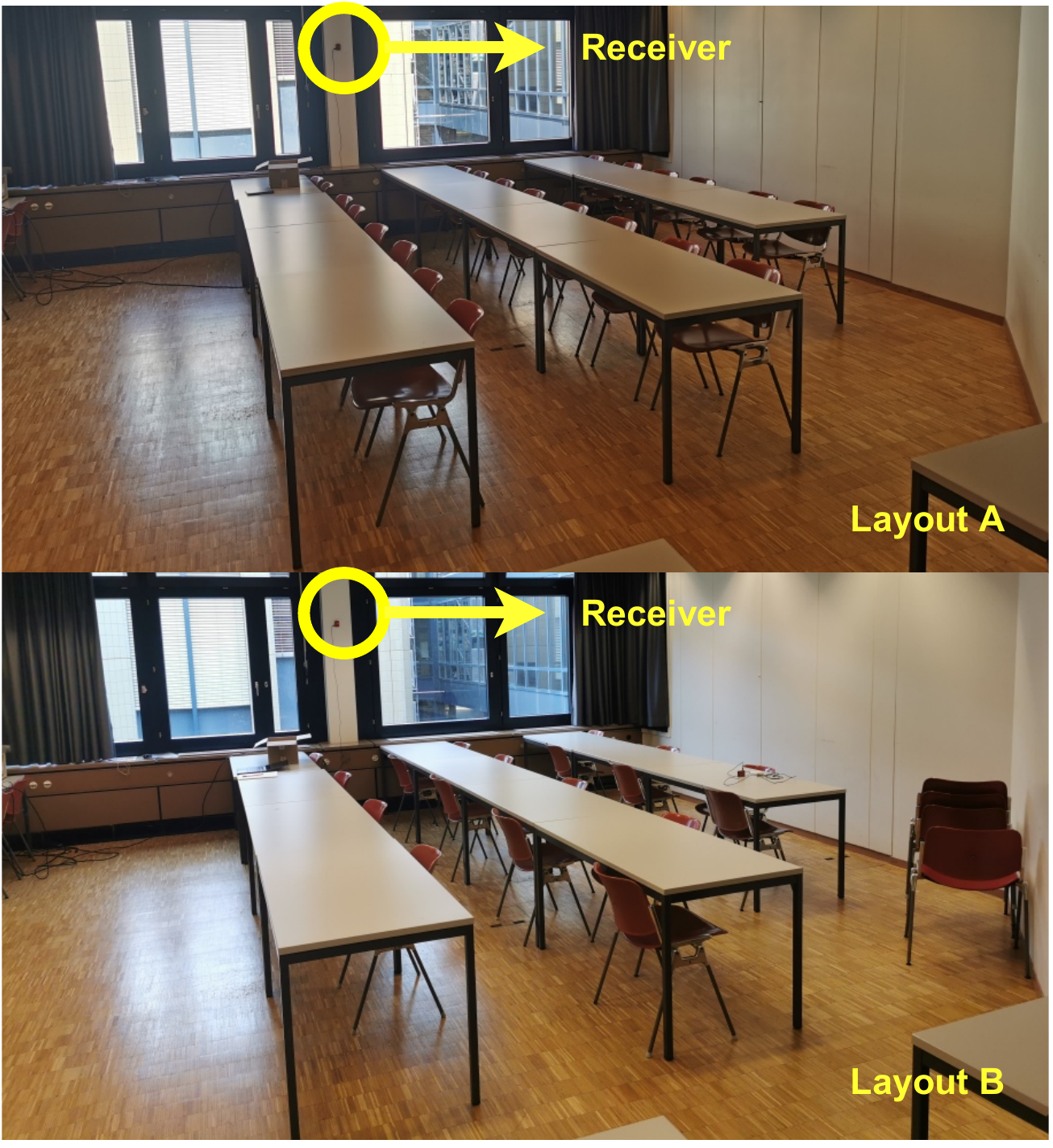}}
    \hfill
  \subfloat[Original computer room layout (top) and layout with moved objects marked with red circles (bottom), the receiver is next to the camera.\label{fig:labroom_photo}]{%
        \includegraphics[width=0.47\linewidth]{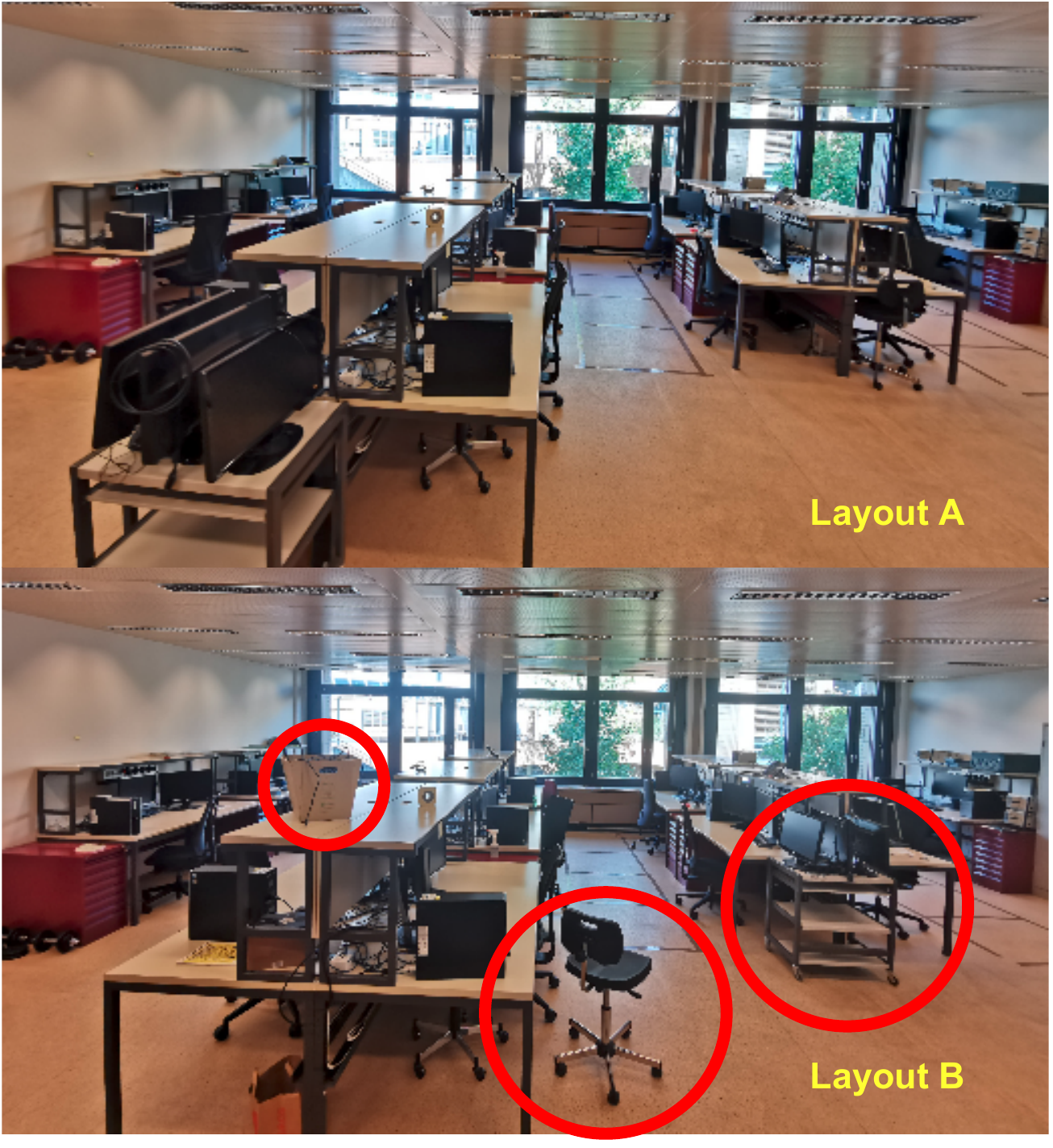}}
  \caption{Layout changes in the large-scale setups}
  \label{fig:changes} 
  \vspace{-0.4cm}
\end{figure}

\begin{table*}[]
\centering
\resizebox{\textwidth}{!}{
\begin{tabular}{@{}ccccccccc@{}}
\toprule
                                                                                     &                               & \textbf{\begin{tabular}[c]{@{}c@{}}Small-scale \\ scenario\end{tabular}} & \textbf{\begin{tabular}[c]{@{}c@{}}Classroom \\ A1$\rightarrow$A2\end{tabular}} & \textbf{\begin{tabular}[c]{@{}c@{}}Classroom \\ A1$\rightarrow$B\end{tabular}} & \textbf{\begin{tabular}[c]{@{}c@{}}Computer \\ room (LOS)\\ A1$\rightarrow$A2\end{tabular}} & \textbf{\begin{tabular}[c]{@{}c@{}}Computer \\ room (LOS)\\ A1$\rightarrow$B\end{tabular}} & \textbf{\begin{tabular}[c]{@{}c@{}}Computer \\ room (NLOS)\\ A1$\rightarrow$A2\end{tabular}} & \textbf{\begin{tabular}[c]{@{}c@{}}Computer \\ room (NLOS)\\ A1$\rightarrow$B\end{tabular}} \\ \midrule
\multicolumn{1}{c}{\multirow{2}{*}{\textbf{1D-GMM~\cite{mohammadmoradi_uwb-based_2019}}}}                                & \multicolumn{1}{c}{1 snapshot}   & \multicolumn{1}{c}{61.1\%}                                            & \multicolumn{1}{c}{41.1\%}                                            & \multicolumn{1}{c}{33.8\%}                                            & \multicolumn{1}{c}{47.1\%}                                                       & \multicolumn{1}{c}{41.4\%}                                                      & \multicolumn{1}{c}{28.1\%}                                                        & 26.2\%                                                                            \\ \cmidrule(l){2-9} 
\multicolumn{1}{c}{}                                                                & \multicolumn{1}{c}{MV100}    & \multicolumn{1}{c}{77.2\%}                                            & \multicolumn{1}{c}{67.5\%}                                            & \multicolumn{1}{c}{56.6\%}                                            & \multicolumn{1}{c}{87.9\%}                                                       & \multicolumn{1}{c}{76.5\%}                                                      & \multicolumn{1}{c}{56.5\%}                                                        & 50.6\%                                                                            \\ \midrule
\multicolumn{1}{c}{\multirow{2}{*}{\textbf{MD-GMM}}}                                & \multicolumn{1}{c}{1 snapshot}   & \multicolumn{1}{c}{92.5\%}                                            & \multicolumn{1}{c}{59.8\%}                                            & \multicolumn{1}{c}{49.4\%}                                            & \multicolumn{1}{c}{56.6\%}                                                       & \multicolumn{1}{c}{44.9\%}                                                      & \multicolumn{1}{c}{37.0\%}                                                        & 33.8\%                                                                            \\ \cmidrule(l){2-9} 
\multicolumn{1}{c}{}                                                                & \multicolumn{1}{c}{MV100}    & \multicolumn{1}{c}{99.4\%}                                            & \multicolumn{1}{c}{99.9\%}                                            & \multicolumn{1}{c}{94.8\%}                                            & \multicolumn{1}{c}{97.7\%}                                                       & \multicolumn{1}{c}{85.7\%}                                                      & \multicolumn{1}{c}{86.2\%}                                                        & 81.2\%                                                                           \\ \midrule
\multicolumn{1}{c}{\textbf{\begin{tabular}[c]{@{}c@{}}MD-GMM \\-MaxSim\end{tabular}}} & \multicolumn{1}{c}{100 snapshots} & \multicolumn{1}{c}{99.8\%}                                            & \multicolumn{1}{c}{100.0\%}                                            & \multicolumn{1}{c}{94.9\%}                                            & \multicolumn{1}{c}{98.5\%}                                                       & \multicolumn{1}{c}{88.9\%}                                                      & \multicolumn{1}{c}{92.5\%}                                                        & 90.1\%                                                                            \\ \midrule
\multicolumn{1}{c}{\textbf{\begin{tabular}[c]{@{}c@{}}MD-GMM \\-SVC\end{tabular}}} & \multicolumn{1}{c}{100 snapshots} & \multicolumn{1}{c}{99.7\%}                                            & \multicolumn{1}{c}{99.9\%}                                            & \multicolumn{1}{c}{95.7\%}                                            & \multicolumn{1}{c}{99.4\%}                                                       & \multicolumn{1}{c}{89.8\%}                                                      & \multicolumn{1}{c}{95.0\%}                                                        & 91.9\%                                                                            \\ \bottomrule
\end{tabular}
}
\vspace{0.1cm}
\caption{Localization accuracy with different methods in different scenarios. Majority voting among $100$ snapshots is denoted by MV100.}
\label{tab:test_accuracy}
\vspace{-0.4cm}
\end{table*}
\vspace{-0.25cm}
\subsection{Experiment Setup}
We use the same \gls{uwb} chip DW1000~\cite{decawave_dw1000_nodate} for both the  transmitter and the receiver.
The transmitter is moving in the considered areas, sending packets in \gls{uwb} channel $5$ with an interval of $0.02$\,s.
The receiver records and stores the first $50$ bins in the \gls{cir}. 
Note that in DW1000, the \gls{cir} bins are interpolated such that each bin represents half a period of the $499.2$\,MHz fundamental frequency~\cite{decawave_dw1000_nodate_um}, so we filter the \gls{cir} in the frequency domain and down-sample the \gls{cir} such that we only need to consider $25$ bins after filtering.
We considered the three following measurement scenarios.
\vspace{-0.25cm}
\subsubsection{Small-scale scenario}
First, we test in a small-scale scenario, where the transmitter is attached to a two-axis \gls{cnc} machine and the receiver is located at a distance of $3$\,m.
The \gls{cnc} is programmed to move the transmitter within the pre-defined areas of size $8.5$\,cm by $8.5$\,cm, which also corresponds to the scale of the setup in~\cite{mohammadmoradi_uwb-based_2019}.
\vspace{-0.25cm}
\subsubsection{Large-scale scenario (classroom)}
We also conduct measurements in a classroom of size $5.6$\,m$\times7.8$\,m.
The receiver is located $2$\,m above the floor and the transmitter is carried by a walking person, in positions in which people typically hold their phones.
The receiver is most of the time in the \gls{los} of the transmitter, except for the moment when the carrying person is blocking the \gls{los} depending on the walking direction.
The considered area consists of $12$ sub-areas of size $1.4$\,m$\times 0.95$\,m as shown in Fig.~\ref{fig:classroom}.
In order to test the robustness of the different methods against environment changes, we move the chairs in the classroom as shown in Fig.~\ref{fig:classroom_photo} to create a second room layout and collect two datasets with two different room layouts.
\vspace{-0.25cm}
\subsubsection{Large-scale scenario (computer room)}
We also collect data in a more complex environment that is equipped with desks, shelves, and computers as shown in Fig.~\ref{fig:labroom} of size $6.5$\,m$\times 10$\,m.
The sub-areas are $1$\,m$\times 1$\,m.
The receiver is located $1.8$\,m above the floor.
The computer room allows us to test both \gls{los} and \gls{nlos} scenarios. 
If the transmitter is held in hand, the \gls{los} is most of the time blocked by shelves and computers.
If the transmitter is held high up in the air, the \gls{los} is guaranteed most of the time.
Two datasets are collected with different layouts as shown in Fig.~\ref{fig:labroom_photo}.
\vspace{-0.35cm}
\subsection{Data Collection}
In the small-scale scenario, we test the performance with \gls{los} and without environment changes. 
Both training and test sets contain $5500$ snapshots.
In the large-scale scenarios (classroom, computer room (\gls{los}) and computer room (\gls{nlos})), two datasets are collected for two different layouts and are denoted by set A and set B in each scenario.
For the classroom, set A contains $5200$ snapshots. The first $4000$ snapshots are denoted as training set A1 and the last $1200$ snapshots are denoted as test set A2. 
Set B contains $1200$ snapshots with layout changes compared to set A for test. 
For the computer room case, we collect more data since the environment is more complicated. 
Set A contains $12000$ samples.
The first $10800$ snapshots are denoted as training set A1 and the last $1200$ snapshots are denoted as test set A2. Set B contains $1200$ snapshots for test again with layout changes. 
We test the robustness against the room layout change by training with the samples in set A1, and testing with the samples in set A2 and set B in each scenario.
For the 1D-GMM method and the MD-GMM method, majority voting is performed over $100$ snapshots.
For the similarity metric-based methods (MD-GMM-MaxSim and MD-GMM-SVC), each sample set contains $100$ snapshots for a fair comparison.
\vspace{-0.3cm}
\subsection{Offline Learning Parameters}
In the offline phase, we use the variational Bayesian estimation of a Gaussian mixture function provided by the \textit{scikit-learn} toolbox~\cite{scikit-learn}, with a maximum of $2$ Gaussian components for 1D-GMM and $5$ Gaussian components for MD-GMM, a maximum of $10^{4}$ iterations, a convergence threshold of $10^{-3}$ and a weight concentration prior of $10^{-3}$.
We use the SVC supported by the same toolbox, with the radial basis function kernel and maximum iteration of~$10^{4}$.
The other parameters are as default in the toolbox version~1.0.2.
\vspace{-0.45cm}
\subsection{Results}
The localization accuracy is shown in Table~\ref{tab:test_accuracy}.
Both GMM-based methods benefit from majority voting.
The 1D-GMM shows good results in the small-scale environment with an accuracy of $77.2\%$ with majority voting.
However, when the considered areas become larger, as in the classroom and the computer room, MD-GMM shows a considerably better performance than 1D-GMM. 
After majority voting, the MD-GMM achieves up to $99.9\%$ accuracy in the classroom setup, and is quite robust to room layout change with an accuracy of $94.8\%$. 
In more complicated environments in the computer room, with the \gls{los} scenario, MD-GMM achieves an accuracy of $97.7\%$ without layout changes, and $85.7\%$ with layout changes.
By introducing the similarity metric (MD-GMM-MaxSim and MD-GMM-SVM), the accuracy can be improved slightly by around $2\%$.
However, in an environment with more \gls{nlos} cases, the similarity metric-based methods improve the accuracy by around $6\%$ to $9\%$ compared to MD-GMM with the majority voting.
Even with layout changes, the MD-GMM-MaxSim still shows significant improvement and reaches an accuracy of $90.1\%$ compared to the accuracy of $81.2\%$ by MD-GMM.
The \gls{svc}-based method shows slightly better performance compared to the MaxSim method in the \gls{nlos} scenario and the improvement is around $2\%$.

Regarding online estimation efficiency, estimation with 1D-GMM is about $16$ times slower than with MD-GMM, since 1D-GMM needs to loop over each delay bin to calculate the probability, whereas MD-GMM considers all the delay bins jointly.
Among all MD-GMM-based methods, using the \gls{svc} increases the run time by $20\%$, compared to the MD-GMM-MaxSim and the majority voting method.

\vspace{-0.4cm}
\section{Conclusions}
\vspace{-0.1cm}
In this paper, we showed that learning statistics of multiple \gls{cir}s leads to good performance in \gls{uwb} single-anchor localization in environments with a rich amount of multipath. 
By properly combining the information obtained from multiple \gls{cir} delay bins, the localization accuracy shows significant improvement by up to $30\%$ compared to considering each delay bin separately as done until now in the literature in large-scale environments. 
Instead of combining multiple hard decisions based on the probability of snapshots in \gls{gmm}s of reference areas, directly considering the snapshots as a sample set and comparing the similarity metric between the \gls{cir} set and the reference \gls{cir} sets performs better.
Especially in rooms with complex layouts and many obstacles, the similarity metric-based methods can further improve the performance by up to $10\%$ compared to the majority voting method.

\vfill\pagebreak

\balance
\bibliographystyle{IEEEbib}
\bibliography{refs}

\end{document}